\documentstyle[12pt,epsf]{article}
\def\be{\begin{equation}}
\def\ee{\end{equation}}
\def\ba{\begin{eqnarray}}
\def\ea{\end{eqnarray}}
\def\bq{\begin{quote}}
\def\eq{\end{quote}}

\begin{document}
\thispagestyle{empty}
\begin{flushright}
SU-ITP-98-21\\ hep-th/9804062\\ April 1998
\end{flushright}
\vspace*{1cm}
\begin{center}
{\Large \bf Entropy Count for Extremal Three-Dimensional Black
Strings}\\
\vspace*{2cm}
Nemanja Kaloper\footnote{E-mail: kaloper@leland.stanford.edu}\\
\vspace*{0.2cm}
{\it Department of Physics}\\
{\it Stanford University}\\
{\it Stanford, CA 94305-4060}\\
\vspace{3cm}
ABSTRACT
\end{center}
We compute the entropy of extremal black strings in three
dimensions, using Strominger's approach to relate
the Anti-de-Sitter near-horizon geometry and
the conformal field theory at the asymptotic infinity of this
geometry. The result is identical to the geometric
Bekenstein-Hawking entropy. We further discuss an embedding of
three-dimensional black strings in $N=1~D=10$ supergravity and
demonstrate that the extremal strings preserve $1/4$ of
supersymmetries.
\vfill
\setcounter{page}{0}
\setcounter{footnote}{0}
\newpage

There has been tremendous interest recently in exploring the
proposal that a supergravity theory on an Anti-de-Sitter ($AdS$)
space is equivalent to a superconformal theory on the $AdS$
boundary \cite{mald,gkp,witt}. A strong evidence for this result
has been provided by the connection between the isometries of the
$AdS$ geometry and the conformal theory on the boundary
\cite{renata}. This connection is clearly seen in the context of
black $p$-branes in supergravity theories, where the near-horizon
geometry of a number of extremal supersymmetric $p$-brane systems,
either RR or NSNS or both, is $AdS$ \cite{mald,renata,frons,ferr}.
One therefore believes that there must be a natural projection of
the bulk theory onto the boundary, and that the boundary theory,
like a hologram, retains the information about the bulk degrees of
freedom which were projected down \cite{witt}.

One application of the connection between the bulk supergravity and
the boundary conformal field theory has been in the counting of
microstates which give rise to the black hole entropy. There are
different methods for accomplishing this goal. One approach begins
with establishing a $U$-duality map \cite{hyun}
between higher dimensional black holes and the three-dimensional
$AdS$ Ba\~nados-Teitelboim-Zanelli (BTZ) black hole \cite{btz}. The
BTZ black hole is also a solution of string theory \cite{mien}.
Using $D$-brane method for counting black hole microstates
\cite{vafstr,calmald}, which has been applied with great success in
higher dimensions \cite{rob,hormalstr}, one can show that the
$D$-brane count for the BTZ black hole embedded in type II string
theory precisely matches the Bekenstein-Hawking formula
\cite{hyun,birming}. However, the $D$-brane counting methods are
designed for solutions which can be continued to the weak coupling
regime. It is of interest to see how one can count microstates in
the strong coupling, when the black hole causal structure is
manifest. Using the U-duality-generated trans-dimensional chart of
related solutions \cite{bps}, Sfetsos and Skenderis have reproduced
the Bekenstein-Hawking formula for a number of solutions
\cite{sfeske}, which can be arbitrarily far from extremality.
They have obtained the entropy using Carlip's approach
\cite{carlip} to count the microstates of the BTZ black hole
formulated as a solution of topological $2+1$ gravity.
In this case, one counts the entropy by associating degrees of
freedom to the horizon, which explains why the entropy of the hole
is measured by the horizon area, and not by the volume of the space
excised from the spacetime to accommodate the hole. However this
is based on the description of the BTZ hole by a purely topological
Chern-Simons formulation, which is not a fundamental description of
a sector of string theory.
Another, more direct, method of counting the entropy has been
proposed recently by Strominger \cite{strom}. Similar approach has
also been advocated by Birmingham, Sachs and Sen
\cite{bss}. This method is based on the fact
that if a solution of a dynamical theory of gravity has a region
which is approximated by $AdS_3$ geometry, one can
construct a conformal field theory (CFT)
on the boundary of $AdS_3$.
The boundary theory can be determined by studying the
large distance limit of diffeomorphisms in the
$AdS$ bulk \cite{bh}. The $AdS$ boundary need
not be the physical infinity of the spacetime, but
the limit of validity of the $AdS$ approximation
to the bulk solution. On the $AdS_3$
boundary, the ``large" diffeomorphisms
cease to be integrable, and
acquire local dynamics \cite{rt,bh,carlip,banados}.
Their dynamics is given by a CFT, and one can determine
the degeneracy of a black hole macrostate by counting different
microstates of the CFT which correspond to the same black hole
solution.
This should match the
Bekenstein-Hawking formula. Similar investigations have been
successfully carried out for
the BTZ black hole \cite{strom,bss}, the whole of
de-Sitter space \cite{stromald}, and some 5D (near)extremal black
strings with the near-horizon geometry approximated by a
(near)extremal BTZ hole \cite{ballar}.
For other uses, see also \cite{others}.

In this article, we discuss a simple realization of the boundary
conformal theory for extremal three-dimensional black strings,
constructed by Horne and Horowitz \cite{horhor} prior to the
discovery of the BTZ black hole. This solution in fact corresponds to
the throat limit of a fundamental string inside an NS $5$-brane,
as we will discuss later \cite{fiveb}. Near the horizon, the scalar
dilaton is fixed, approaching a constant value very quickly. Hence
the near-horizon geometry of this solution is $AdS_3$. Therefore, we
can straightforwardly apply Strominger's
procedure \cite{strom}, up to
the distances of order of $\sqrt{Ml}$
away from the horizon, where
$M$ corresponds to the mass per unit length of the string.
The near-horizon limit of the dilaton and
the NSNS $3$-form specify the effective Newton's constant and the
cosmological constant, which in turn define the CFT central charge
and normalization. This suffices to specify the
Virasoro algebra of the
boundary CFT. In order to find the degeneracy of the string solution
from the boundary CFT, we use the fact that this
Virasoro algebra is faithfully
reproduced by the boundary CFT of the
topological Chern-Simons $SO(2,2)$
gravity. The boundary degrees of freedom are in one-to-one
correspondence with the large gauge transformations in the bulk,
and we can compute the number of states using the
microcanonical ensemble
approach of \cite{bbo}. In the course
of our calculation, we encounter
an interesting subtlety: we must regularize the expressions for the
boundary CFT. The proper regularization of the theory consists of
considering a near-extremal solution, and compactifying the length
along the string with a period inversely proportional to the
Hawking temperature. This reproduces the correct expression
for the statistical entropy, which matches exactly the
Bekenstein-Hawking formula for the extremal black string. Finally,
we show how black strings arise as consistent supersymmetric
solutions preserving $1/4$ of supersymmetries of $N=1~D=10$
supergravity.

We first review here the black string family of \cite{horhor}, with
the attention on the entropy. To construct a $3D$ black string
solution, we look for the static extrema of the low-energy action
\be
I_3 = \frac{1}{16\pi G}\int d^3 x \sqrt{g} e^{-\phi}
\Bigl(R + (\nabla \phi)^2 - \frac{1}{12}
H^2_{\mu\nu\lambda} + \frac{4}{l^2} \Bigr)
\label{3deff}
\ee
where in addition to the metric $g_{\mu\nu}$ we have the dilaton
$\phi$ and the NSNS $2$-form field strength $H_{\mu\nu\lambda} =
\partial_\mu B_{\nu\lambda} + cyclic ~ permutation$. The term
$\Lambda = - 2/l^2$ plays role of the effective (negative)
stringy cosmological constant, and can
arise if the central charge of the three-dimensional theory is not
equal to the dimension of the spacetime.
In the Wess-Zumino-Witten construction of \cite{horhor}, it is given
by the level of the Ka\v c-Moody algebra $l^2 = k/2$.
The cosmological term can also
arise from dimensional reductions on internal manifolds with
nonabelian symmetry. We will give an example later.

The static black string family which emerges from (\ref{3deff}) is
given by
\ba
ds^2 &=& - \bigl(1-\frac{M}r\bigr) dt^2 +
\bigl(1-\frac{Q^2}{Mr}\bigr) dx^2 +
\frac{l^2 dr^2}{4 (r - M)(r-Q^2/M)} \nonumber \\
&&H = \frac{Q}{r^2} ~dt \wedge dx \wedge dr ~~~~~~~~~~~~~~~~~
e^{-\phi} = ~\frac{r}{l}
\label{blssol}
\ea
The parameters $M$ and $Q$ correspond to the ADM mass
per unit length of the string, and
the axion charge per unit length, respectively.
They are related to the mass $m$ and spin $j$
of the dual BTZ black hole, in the notation of \cite{strom},
according to $M = 4G(ml + \sqrt{m^2l^2-j^2})$
and $Q=4Gj$, and have the dimension of length.
We have normalized the solution such that the case
$M=Q=0$ corresponds to the linear dilaton solution of \cite{ldv},
which is the black string vacuum.
This solution is asymptotically flat
(in the string frame) as $r \rightarrow \infty$, has an event horizon
at $r_+ = M$, a Cauchy horizon at $r_-=Q^2/M$, and a curvature
singularity at $r \rightarrow 0$. The singularity is reached by
all future-oriented null geodesics, and hence the manifold is
geodesically incomplete. There is a subtlety in defining the
complete Penrose diagram of this solution. The diagram should be
three-dimensional to accommodate the interplay between the
coordinates $t$ and $x$, and the details can be found in
\cite{horhor}.

A black string emits Hawking radiation to ${\cal J}^+$, and
looses energy while its charge is conserved because of the Gauss
law for $H$. The dynamics of emission is that of a black body with
the Hawking temperature $T_H$, which can be determined after a
simple Wick rotation $t \rightarrow i \tau$. Near the event
horizon, $r\rightarrow r_+ = M$, we recover the Euclidean metric in
cylindrical polar coordinates, which is smooth near the origin if
we identify the Euclidean time with
the period $2 \pi l M/(M^2 - Q^2)$.
Hence the Hawking temperature of the
black string is
\be
T_H = \frac{\sqrt{M^2 - Q^2}}{2~\pi lM}
\label{hawktemp}
\ee

Clearly, the temperature vanishes when $M$ reduces to $Q$, and the
resulting solution ceases to emit Hawking radiation. This
corresponds to the extremal limit of the black string.
The horizon is still at $r_+ = M$. It is tempting to think
that the analytic continuation across the horizon is accomplished
by going to $r < M$. However, this is not true \cite{horhor}. The
coordinate $r$ cannot be continued across the horizon, which is a
fixed point for $r$. But this is only an artifact of a bad
coordinate system. The correct continuation is performed with
choosing the new radial coordinate $\rho$ according to $r = M +
\rho^2/l$, and letting $\rho$ pass through zero, which is the
location of the horizon in the new coordinate system \cite{horhor}.
In terms of the new radial coordinate, the solution is
\ba
&&~~~~~ds^2 = \frac{\rho^2}{M l + \rho^2}
\bigl(-dt^2 + dx^2\bigr)
+ \frac{l^2 d\rho^2}{\rho^2} \nonumber \\
H &=& \frac{2Ml\rho}{(M l +
\rho^2)^2} ~dt \wedge dx \wedge d\rho ~~~~~~~~~~~~~
e^{-\phi} = \frac{Ml + \rho^2}{l^2}
\label{blssolextrad}
\ea
The solution is invariant under $\rho \leftrightarrow - \rho$,
which implies that the universe behind the event horizon is a
mirror image of the original one. Hence the maximal extension of
this solution is an infinite array of asymptotically Minkowski
diamonds aligned along a zig-zag event horizon, as mentioned
before. This is illustrated in Figure 1.

The Bekenstein-Hawking entropy is in
general given as one quarter of the area of event horizon,
$S_{BH} = {\cal A}/(4 G_N)$, in Planck units. This formula is
expected to work well for sufficiently large configurations,
where we can trust the classical theory close to the event
horizon. The normalization of the entropy is given in the units set
by the normalization of the Einstein term in the action,
$\sim \int d^D x \sqrt{g} R/(16\pi G_N)$.
For the black string in the
string frame, the Newton's constant is $G_N = G \exp(\phi(r_+))$,
as can be verified from (\ref{3deff}).
The ``area" in this case is just the proper length of the string.
It is given by ${\cal A} = \int \sqrt{g_{xx}(r_+)} dx
= \sqrt{M^2 - Q^2} L/M$, where
$L = \int dx$ is the comoving length of the string. Hence
substituting these formulas, we find
\be
S_{BH} = \frac{{\cal A}}{4G} e^{-\phi(r_+)}
= \frac{L}{4Gl} \sqrt{M^2 - Q^2}
\label{entL}
\ee
For fixed $M$ and $Q$, this expression diverges as $L \rightarrow
\infty$, which corresponds to an infinite string. However, the
entropy per unit length is given by
$S_{BH}/L = \sqrt{M^2 - Q^2}/(4Gl)
= \pi M T_H/(2G)$, and it is finite (and vanishing
when $T_H = 0$). This suggests that in order to consider the
entropy of the string we can break up the whole string into an
infinite array of identical ``elementary strings", each of them
being a $T$-dual image of a BTZ black hole. In effect,
we formally compactify the coordinate along the string on a
circle of circumference $L$, and then view the infinite string as
the covering space of an elementary segment \cite{hw}.
The length of the
elementary segment can be determined by comparing it to the
$T$-dual BTZ hole. Following \cite{hw}, after straightforward
algebra we find it to be
\be
L = 2 \pi l \sqrt{\frac{Ml}{M^2 - Q^2}}
\label{Lcomp}
\ee
Substituting this expression in the formula for the
Bekenstein-Hawking entropy, we find that the entropy of each
elementary string segment is
\be
S_{BH} = \frac{\pi}{2 G} \sqrt{Ml}
\label{bekhaw}
\ee

Using the picture where the string is broken into
elementary segments ensures that
the formula (\ref{bekhaw}) for the Bekenstein-Hawking
entropy is valid in the extremal limit.
This may sound a little odd, since as we have seen above,
the entropy per unit length of the string is directly proportional
to the Hawking temperature of the string, and hence zero in the
extremal limit. However, the length  $L$ of the elementary
segments of a string diverges as $M \rightarrow Q$. In the
extremal limit, the whole infinite string should be viewed as a
single elementary segment itself. This is deduced from the fact
that this extremal string is dual to the extremal spinning BTZ
black hole \cite{mien,hw}.
A spinning black hole in turn can be obtained by
boosting a static hole along the circle and identifying the
boosted coordinate. The extremal limit corresponds to the
infinite Lorentz boost, and the fact that the dual extremal string
must be infinite arises from the need to compensate for the
(inverse) Lorentz-contraction of the azimuthal angle of the BTZ
hole. The nonvanishing total entropy
$S_{BH} = \frac{\pi}{2G} \sqrt{Ml} $
can be viewed as the zero point entropy
of the extremal black string.

Now we can proceed to compute the statistical entropy of the black
string system, using the proposal of \cite{strom}. This should be
applicable, since the near-horizon geometry of the extremal black
string is approximately $AdS$. We first sketch the calculation, and
point out a subtlety which arises in its course. Using the solution
(\ref{blssolextrad}), we see that as $\rho \rightarrow 0$, the
metric approaches
\be
ds^2 = \frac{\rho^2}{Ml} \bigl(-dt^2 + dx^2\bigr) + \frac{l^2
d\rho^2}{\rho^2}
\label{nearh}
\ee
which is clearly $AdS$. In this limit, $H^2 = -\frac{24}{l^2}
\frac{M^2 l^2}{(Ml+\rho^2)^2} \rightarrow
-\frac{24}{l^2}$ and also $(\nabla \phi)^2
= \frac{4\rho^4}{l^2(Ml+\rho^2)^2}
\rightarrow 0 $.
The effective action is approximated by
\be
S_{eff} = \frac{M}{16\pi Gl} \int d^3x \sqrt{g} \Bigl(R +
\frac{2}{l^2} \Bigr)
\label{approxact}
\ee
The effective cosmological constant is half of the value of the
stringy cosmological constant $2/l^2$ due to the contribution of the
$3$-form field strength:
$1/l^2_{eff} = 2/l^2 + H^2/24 = 2/l^2 - 1/l^2
= 1/l^2$. The sign in the first equation is plus,
because the solution for $H$ must be inserted into the action with
care, using the Lagrange multiplier
method \cite{tmg}. Hence, near the
horizon, the only effect of the dilaton and the $3$-form is to
renormalize the Newton's constant and the cosmological constant of
the effective theory, which become
\be
G_N = \frac{Gl}{M}~~~~~~~~~~~ \Lambda_{eff} = \frac{1}{l^2}
\label{consts}
\ee
The remainder of the counting procedure is to find the boundary
CFT and count the density of states.
According to \cite{strom,bss}, the
boundary CFT is given by the projections of the
``large" diffeomorphisms of the bulk
theory at asymptotic infinity.
A black string macrostate is specified by the
$SL(2,R)$ sector of the boundary Virasoro algebra.
This information is encoded in the long-distance behavior
of the dreibein and the
spin connexion. To obtain it, we resort to the
Chern-Simons gauge theory formulation of three-dimensional
gravity, with the same Newton's constant and cosmological constant.
This theory produces a faithful representation of the
boundary Virasoro algebra, which is
homomorphic with the string boundary CFT, since both are given
by the ``large" diffeomorphisms. We will use
the microcanonical ensemble
approach of \cite{bbo}
to find the degeneracy of the boundary theory.

If we straightforwardly proceed in this direction, we encounter an
undetermined expression for the entropy of the type $0 \cdot
\infty$. This ambiguity arises because in the
extremal limit the Hawking temperature vanishes, while the circle
spanned by the $x$ coordinate decompactifies. We have encountered
the same problem earlier, while determining the Bekenstein-Hawking
entropy of the extremal solution. Indeed, in the statistical
approach, as \cite{bbo} show, the expressions for the number of
chiral states on the boundary involve integrations over $x$, which
diverges as $M \rightarrow Q$. This is multiplied by the number
density of the states on the boundary, which is essentially given
by the mass and the spin of the metric in the $AdS$ limit. By
comparing the metric (\ref{nearh}) to the BTZ solution in this
limit, we see that in the $AdS$ limit the solution
(\ref{blssolextrad}) is approximated by the $M=J=0$ BTZ hole.
Naively, this suggests that the number density of the states on the
boundary vanishes. Ergo, we end up with $S \sim 0 \cdot \infty$.

Hence we must first regularize the solution. We do it by
considering near-extremal solutions, with $M^2 = Q^2 +
\epsilon^2$, in the near-horizon limit.
Thus we have $Q^2 = M^2 - \epsilon^2$, and we keep $\rho^2 = r -
M$. The metric becomes
\be
ds^2 = - \frac{\rho^2}{Ml+\rho^2} dt^2 +
\frac{\rho^2 + l\epsilon^2/M}{Ml + \rho^2} dx^2
+ \frac{l^2 d \rho^2}{\rho^2 + l\epsilon^2/M}
\label{appmet}
\ee
The $3$-form and the dilaton still give
$H^2 \rightarrow - \frac{24}{l^2}$,
$(\nabla \phi)^2 \rightarrow 0$ in the limit
$\rho, \epsilon \rightarrow 0$, implying that our near-horizon
approximation is valid up to distances $\rho \sim \sqrt{Ml}$.
To define the boundary CFT, we recall that the effective action is
still given by (\ref{approxact}), with the same values of the
Newton's and cosmological constants as before. The boundary CFT
algebra corresponds to the ``large" diffeomorphisms of this theory.
We recall that the only effect of the matter is to renormalize the
``bare" cosmological constant and Newton's constant. Since
the bulk is $AdS_3$, its symmetry group is
$SO(2,2) \sim SL(2,R) \times SL(2,R)$. Hence the boundary
algebra splits into a direct product of two disjoint
copies of the Virasoro
algebra \cite{bh}. When the boundary is compactified to a
cylinder, which is the case of our
elementary segments, we can write the Virasoro algebras in
terms of the Fourier modes of the diffeomorphism
generators on the boundary:
\be
[L_m, L_n] = i (m-n) L_{m+n} + i \frac{c}{12} (m^3 - m)
\delta_{m+n,0}
\label{vira}
\ee
each with the central charge
\be
c = \frac{3 l_{eff}}{2 G_N}
\label{central}
\ee
For the values of $G_N$ and $l_{eff}$ in
(\ref{consts}), the central charge is $c = \frac{3M}{2G}$.
In order to
determine the degeneracy of the black string, which corresponds to
its entropy, we need to evaluate the eigenvalues of the operators
$L_0$ and $\bar L_0$ in the background (\ref{appmet}). For the BTZ
hole, this leads to simple expressions where $L_0 =
Ml + J$ and $\bar L_0 = M l - J$. But as we have pointed out above,
this would lead to the
incorrect expressions $L_0 = \bar L_0 = 0$ for the case of the
extremal string. We can correctly determine $L_0$ and $\bar L_0$
using the microcanonical ensemble method of \cite{bbo} instead.
As we have noted above, according to \cite{strom},
we should consider the large distance limit
of diffeomorphisms of the bulk
gravity near the horizon (\ref{approxact}).
After choosing the proper gauge, these give
rise to local degrees of freedom
on the boundary that determine
the boundary CFT. Given the near-horizon
specification of the geometry in
terms of the mass and charges, we
can count different microstates
that produce the same mass and charges
to find the entropy. For this purpose,
we do not need the exact construction
of the boundary CFT, but merely a
faithful representation of its Virasoro
algebras. To this end, we can consider the
topological Chern-Simons gauge
theory on $SO(2,2)$ \cite{wittop}.
This theory is the effective
description of only the diffeomorphism
sector, since it contains no local degrees of freedom.
Since $SO(2,2) \sim
SL(2,R) \times \bar{SL}(2,R)$,
one can rewrite the action
(\ref{approxact}) as a linear
combination of two copies of the
Chern-Simons gauge theory on $SL(2,R)$.
The tangent space Hodge
dual of the spin connection is $\omega^a = \frac12
\epsilon^{abc} \omega_{bc}$, and the gauge fields are
$A^a_{\pm} = (\omega^a \pm
\frac{1}{l} e^a) J_a$,
where $J^a$ are the $SL(2,R)$ generators, given in terms of the
Pauli spin matrices as $J_0 = (i/2) \sigma_2$, $J_1 = (1/2)
\sigma_3$ and $J_2 = (1/2) \sigma_1$. The inner product on the
algebra is $<J_a, J_b> = Tr(J_a J_b)
= (1/2) \eta_{ab}$, where $\eta_{ab}$
is the tangent space metric. The dreibein $e^a$ and the dual spin
connection $\omega^a$ are treated as independent variables, and the
structure equation $de^a + \epsilon^{abc} e_b \wedge \omega_c = 0$
is obtained as the $\omega_a$ equation of motion. The Chern-Simons
action which yields the same equations of motion as
(\ref{approxact}), after the $3D$ metric is defined by
$g_{\mu\nu} = \eta_{ab} e^a{}_\mu e^b{}_\nu$, is \footnote{Our
conventions as compared to \cite{bbo} differ in the definition of
$k$: $k_{bbo} = -l_{eff}/(4G_N) = - M/(4G)$ in our notation.
Also bear in mind that
$A d A = \epsilon^{\mu\nu\lambda} A_\mu \partial_\nu A_\lambda d^3
x$ etc. Thus, $I_{\pm} = \frac{M}{16\pi G} \int d^3 x
\epsilon^{\mu\nu\lambda}
\eta_{ab} A^a{}_\mu
\partial_\nu A^b{}_\lambda + ...$.}
\be
I_{eff} = I_+ - I_-
\label{twoact}
\ee
where the Chern-Simons actions for the left and right chiral gauge
fields are
\be
I_{\pm} = \frac{M}{16\pi G} \int <A_{\pm} \wedge dA_{\pm} + \frac23
A_{\pm} \wedge A_{\pm} \wedge A_{\pm}>
\label{csact}
\ee

The bulk equations of motion
for the Chern-Simons gauge theory are $F^a_{\pm} = 0$, i.e. the
gauge fields in the bulk are pure gauge, $A^a_{\pm} = 2 \eta^{ab}
Tr( J_b g_\pm^{-1} d g_\pm)$. We can calculate the gauge fields as
follows. Because we are considering the near-extremal string, the
boundary is a simple tensor product $R \times S^1$, where the
circle is parameterized by $x$. It is instructive to normalize the
coordinate along the circle to run from $0$ to $2 \pi$. This is
assured by defining the angle $\theta = 2 \pi x/L$, where $L$ has
been given in (\ref{Lcomp}). Changing the variable $\rho$
to $\rho = \sqrt{l/M} \epsilon \sinh(\vartheta)$, with some simple
algebra, we find the dreibein and the dual spin connexion at an
arbitrary value of $\vartheta$ and in the limit $\epsilon = 0$
(where we have also rescaled the time $t =\sqrt{M/l}(L/2\pi)\tau$
to the Wick-rotated Euclidean time):
\ba
&&e^0 = l \sinh(\vartheta)d\tau ~~~~~~~~~~~~~~~
e^1 = l d \vartheta ~~~~~~~~
e^2 = \frac{l^{3/2}}{\sqrt{M}} \cosh(\vartheta) d\theta \nonumber \\
&&\omega^0 = \sqrt{\frac{l}{M}} \sinh(\vartheta) d\theta
~~~~~~~~~ \omega^1 = 0 ~~~~~~~~~~~
\omega^2 = \cosh(\vartheta) d\tau
\label{drei}
\ea
Using now $A^a_{\pm} = \omega^a
\pm \frac{1}{l} e^a$, we find the gauge fields
\be
A^0_{\pm} = \sinh(\vartheta)
\frac{\sqrt{l} d\theta \pm \sqrt{M} d\tau}{\sqrt{M}}
~~~~~~ A^1_{\pm} = \pm d \vartheta ~~~~~~ A^2_{\pm} =
\cosh(\vartheta)
\frac{\sqrt{M} d\tau \pm \sqrt{l} d\theta}{\sqrt{M}}
\label{gaub}
\ee
The Lie-algebra valued $1$-form gauge field is
\be
A_{\pm}~=~\frac12 \pmatrix{
\pm d\vartheta& e^{\mp \vartheta}
\frac{\sqrt{M} d\tau \pm \sqrt{l} d\theta}{\sqrt{M}}
\cr e^{\pm \vartheta}
\frac{\sqrt{M} d\tau \pm \sqrt{l} d\theta}{\sqrt{M}}
& \mp d\vartheta \cr}
\label{gfmatr}
\ee
Now we must choose the gauge as $\vartheta \rightarrow \infty$.
This is best accomplished by choosing the radial components $A_{\pm
\vartheta} = g^{-1}_{\pm} \partial_{\vartheta} g_{\pm} =
\frac12 ~diag(\pm 1, \mp1)$, from which we find the matrix
\be
g_{\pm}~=~\pmatrix{ e^{\pm \vartheta/2}&0 \cr 0& e^{\mp
\vartheta/2}
\cr}
\label{gtmatr}
\ee
We can use this matrix to transform to the gauge $A_{\pm
~a}(t,\theta) = g_{\pm} A_{\pm~a}(\vartheta, \tau, \theta)
g^{-1}_{\pm}$. In this way, we project out the radial dependence
and obtain the CFT representation at any $\vartheta$.
In particular, we find $A_{\pm \theta}(\theta) = \pm
\sqrt{\frac{l}{M}} J_2$. Hence, $A^2_{\pm \theta} =
\pm \sqrt{l/M}$. These gauge fields can be used to
construct the boundary Virasoro charges. Using our
notation and the definitions of \cite{bbo}
we find
\be
L_0 = \frac{M}{32\pi G}
\int \eta_{ab} A^a_{+\theta} ~A^b_{+\theta} d\theta =
\frac{l}{16 G}
\label{lo}
\ee
and, similarly, also $\bar L_0 = l/(16G)$. Here we have dropped
an irrelevant constant term, since we normalize the entropy
to be zero in the linear dilaton vacuum. Since the Cardy
formula for degeneracy of states of a conformal field theory with
central charge $c$ and excitations $n_L = L_0$ and $n_R = \bar L_0$
\cite{cardy} is valid for $1 << c << n_L + n_R$, we must
have $G << M << l/8$. But this is precisely the
classical limit of the black string, where
we trust the solution. Hence we have
\be
S = 2 \pi \sqrt{\frac{c}{6}}
\bigl(\sqrt{n_L} + \sqrt{n_R}\bigr)
\label{card}
\ee
We find, substituting our expressions,
\be
S = \frac{\pi}{2G} \sqrt{Ml}
\label{result}
\ee
i.e. precisely the Bekenstein-Hawking formula. Hence we see that
the counting of the conformal degrees of freedom exactly reproduces
the geometric entropy of the extremal black string.

Originally, black string solutions in three dimensions have been
considered as solutions of the bosonic string theory, where they have
been implemented as exact gauged $SL(2,R) \times R/R$
Wess-Zumino-Witten models. However such black strings arise in any
superstring theory. They correspond to the throat limit of a fundamental
string inside an NS $5$-brane \cite{fiveb}.
They can be represented as compactifications on
a three-sphere of $N=1~D=10$ supergravity theory, which is a
consistent low energy truncation of any superstring theory. If we
consider only the NSNS fields, in any $10D$ superstring theory we
have
\be
I =\int d^{10}x \sqrt{\hat g} e^{-\hat \phi} \Bigl(\hat R + (\hat
\nabla \hat \phi)^2 - {1 \over 12} \hat H^2 \Bigr)
\label{act10s}
\ee
where the caret denotes $10D$ quantities. Supersymmetry of
toroidal compactifications of this theory to three dimensions
has been studied in \cite{ms,bak}. We can now consider compactifying
this action to three dimensions in a different way,
assuming that the manifold splits
as $M_{10} = M_3 \times S^3 \times T^4$, and that the internal
$4$-torus is trivial, i.e. that it decouples. Setting all its
internal moduli fields to $1$, for simplicity, we see that the
$10D$ and $6D$ fields are identical. Now, we can reduce further
on an $S^3$. We take
\be
d\hat s^2 = g_{\mu\nu} dx^\mu dx^\nu + r^2(x) d\Omega_3
\label{met6}
\ee
The scalar field $r(x)$ is the radius of the internal sphere. We
will require that the NSNS $3$-form field is nonzero on the $S^3$,
given by the monopole configuration
\be
H_{abc} = \sqrt{2}Q \sqrt{\Omega_3} \epsilon_{abc}
\label{mon}
\ee
such that $H^2/12 = Q^2/r^6$.
When we reduce the dilaton to $2+1$ dimensions, we find
$\exp(-\phi) = r^3 \exp(-\hat \phi)$. Further, the Ricci scalar
upon reduction produces an additional ``scalar potential"
term $\sim 1/r^2$. The reduced $3D$ action is
\be
I_3 = \int d^3x \sqrt{g} e^{-\phi}\Bigl(R + (\nabla \phi)^2
 - \frac{1}{12} H^2_{\mu\nu\lambda}
-3\frac{(\nabla r)^2}{r^2}
+ \frac{6}{r^2} - \frac{Q^2}{r^6}\Bigr)
\label{redact}
\ee
This expression arises since the $6D$ Ricci curvature splits into
$\hat R = R + 6/r^2 - 6 \nabla^2 r/r - 6 (\nabla r/r)^2$,
where covariant derivatives are taken with respect to the $3D$
spacetime metric. The $\sim \nabla^2 r$ term is integrated by
parts, and the action is, up to a boundary term, given by
(\ref{redact}).

We choose the parameters $Q,r_0$ such that there exist solutions
with $r=r_0=const$. In this way, we recover the action
(\ref{3deff}), where supersymmetry is inherited from $N=1
~D=10$ supergravity. The condition relating the radius of the $S^3$
and the monopole charge is $r^2_0 = |Q|/\sqrt{2}$.
This theory is equal to our original action (\ref{3deff}) in
three dimensions if we choose $r_0 =\sqrt{2}l$.
This gives us the quantization condition for the radius
$r_0$, if we consider the black string as a WZW model.
We can write the extremal string solution as an exact $10D$
configuration
\ba
&&~~d\hat s^2 = \frac{\rho^2}{Ml +\rho^2}(-dt^2 + dx^2)
+ \frac{l^2 d\rho^2}{\rho^2} + 2l^2 d\Omega_3 + d\vec y^2_4
\nonumber \\
\hat H &=& 2\frac{M l\rho}{(Ml + \rho^2)^2}
dt \wedge dx \wedge d\rho + 4l^2 \sqrt{\Omega_3} ~\vec V_3 ~~~~~~~~
e^{-\hat \phi} = \frac{Ml + \rho^2}{l^2}
\label{6Dstr}
\ea
which is clearly very similar to the self-dual dyonic string in
$6D$ \cite{dfkr} lifted to $10D$. An important difference is that
the ``harmonic" form multiplying the transversal section of the
metric has been shifted by a $-1$ (using the standard terminology
where $h = 1 + q/\vec r^{D-3}$), similar to the
$U$-duals of other string
solutions \cite{hyun,bps}.
If we Hodge-dualize the magnetic part of the $3$-form to a $7$-form,
we can recognize the $5$-brane contribution, to which the solution
reduces exactly when $M=0$. Hence $M$ measures the charge of the
fundamental string, as is clear from the electric part of the $3$-form.

We will now show that our solution preserves $1/4$ of
supersymmetries of $N=1~D=10$ supergravity, similarly to the
selfdual dyonic string \cite{dfkr}. We can see this as follows. The
string-frame supersymmetry transformation rules for the $N=1~D=10$
fermions are \cite{dkl}
\ba
&&\delta \lambda = - \frac{1}{4 \sqrt{2}}
\Gamma^M \partial_M \hat \phi \zeta
+ \frac{1}{24\sqrt{2}} \Gamma^{MNK}
\hat H_{MNK} \zeta \nonumber \\
&&~~~~~~~\delta \psi_M = \partial_M \zeta + \frac14
\Omega_{-~M}{}^{NK} \Gamma_{NK} \zeta
\label{ferms}
\ea
The indices $\{M,N,K\}$ run over the $10D$ tangent space, and the
quantity $\Omega_{-~M}{}^{NK} = \hat \omega_{M}{}^{NK}
- \frac12 \hat H_{M}{}^{NK}$
is the generalized spin connexion. The matrices $\Gamma^M$ comprise
the $10D$ Clifford algebra, and $\Gamma^{M_1 M_2 ... M_k}
= \Gamma^{[M_1} \Gamma^{M_2} ...
\Gamma^{M_k]}$ is their
antisymmetrized product. If there exists a spinor $\zeta$ such that
$\delta \lambda = \delta \psi_M = 0$, the solution possesses
unbroken supersymmetries. Let us determine the conditions which
$\zeta$ must satisfy. Since $\Gamma^M \partial_M \hat \phi
= - \frac{\sqrt{2}}{l}
\frac{\rho^2}{Ml + \rho^2} \Gamma^1$ and
$\frac16 \Gamma^{MNK}
\hat H_{MNK} = \frac{\sqrt{2}}{l}
\frac{Ml}{Ml+\rho^2} \Gamma^1 \Gamma^0 \Gamma^2 +
\frac{\sqrt{2}}{l}\Gamma^3 \Gamma^4 \Gamma^5$, we see that
requiring $\delta \lambda = 0$ gives two conditions on $\zeta$:
\be
(1 - \Gamma^0 \Gamma^2) \zeta = 0 ~~~~~~~~ (1 + \Gamma^1 \Gamma^3
\Gamma^4 \Gamma^5) \zeta = 0
\label{proj}
\ee
The chirality condition of $N=1~D=10$ supergravity is $\Gamma^{11}
\zeta = \zeta$. This means that our two conditions are independent.
Hence our solution selects $1/4$ of the available spinors. We now
turn our attention to $\delta \psi_M$. Since the solution
(\ref{blssolextrad}) is a simple tensor product $M_3 \times S^3$,
the spinors split into a $3D$ part and a spherical part. The
sphere is maximally supersymmetric, and we do not get any new
constraints on $\zeta$ from $\delta \psi_M$ in $S^3$. Then,
assuming that $\zeta$ is independent of the coordinates $t,x$, it
is straightforward to see that as long as $M\in \{0,2\}$, $\delta
\psi_M \sim (1-\Gamma^0 \Gamma^2) \zeta$, which is automatically
zero as long as (\ref{proj}) hold. Finally, we consider $\delta
\psi_1 = 0$. This gives a first order differential equation
\be
\zeta' - \frac{Ml}{4\rho(Ml+\rho^2)}
\Gamma^0 \Gamma^2 \zeta = 0
\label{killsp}
\ee
which, using $\Gamma^0 \Gamma^2 \zeta = \zeta$, is solved by
$\zeta(\rho) = (1/8)
\ln(\frac{\rho^2}{Ml+\rho^2}) \zeta_0$,
where $\zeta_0$ is any $S^3$-symmetric
spinor which solves the algebraic
constraint (\ref{proj}).
Finally we see that there are two
constraints on $\zeta$, and so our solution indeed preserves $1/4$
of supersymmetries. We note that in the vicinity of
the horizon, this solution is $AdS_3 \times S^3$, and it preserves
$1/2$ of supersymmetries, since similarly to \cite{chams,mald}
supersymmetry gets enhanced.

Thus, we have seen that the proposal to count the entropy of
solutions with an $AdS$ near-horizon geometry via the boundary CFT
outside of the horizon reproduces exactly
the Bekenstein-Hawking formula when applied to extremal
three-dimensional black strings. The entropy so obtained should be
identified with the zero-point entropy associated with the string.
It actually corresponds to a vanishing entropy per unit length,
which is consistent with the picture of an infinite string. The
calculation involves a nontrivial step where the solution must be
regularized, by considering the near-extremal case instead. Since
the extremal strings can be embedded in $N=1~D=10$ supergravity,
and preserve $1/4$ of supersymmetries of the theory, the
result should be stable under quantum corrections. An
interesting question which may be asked is if this calculation can
be extended further away from the extremal limit. The
Bekenstein-Hawking formula for the entropy is expressed in terms of
the mass per unit length, and is formally insensitive to how far
from extremality the solution is.
We believe that this deserves further
investigation.

\vspace{1cm}
{\bf Acknowledgements}

I wish to thank R. Kallosh, R.C. Myers and J. Rahmfeld for useful
discussions. I would like to thank R.C. Myers for very
valuable comments on the manuscript.
This work has been supported in part by
NSF Grant PHY-9219345.

\newpage

\begin{center}
\begin{figure}
\leavevmode
\hbox{\epsfysize=12cm {\epsffile{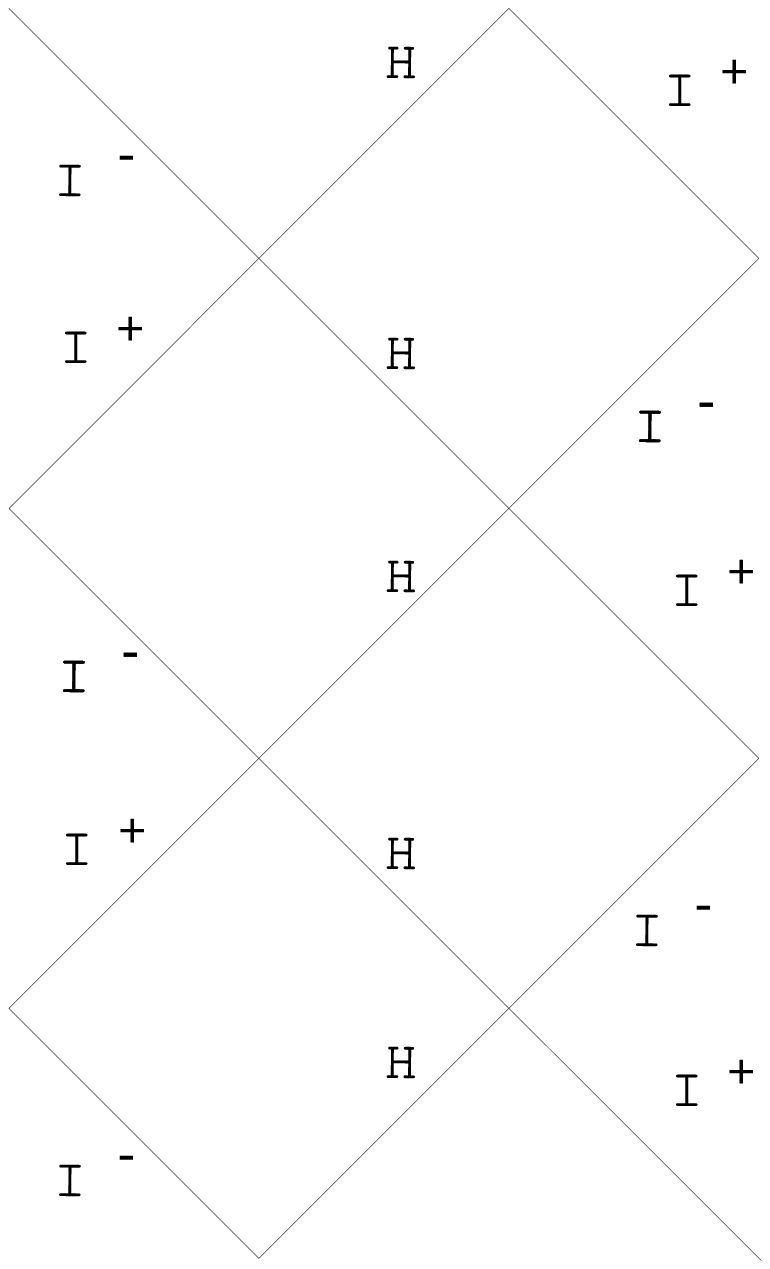}}}
\caption{Maximal extension of the extremal
Horne-Horowitz black string. H denotes
the event horizon, and I$^+$
and I$^-$ future and past null infinity, respectively.}
\end{figure}
\end{center}

\end{document}